# QDEE: Question Difficulty and Expertise Estimation in Community Question Answering Sites

**Jiankai Sun**
The Ohio State University
sun.1306@osu.edu

**Sobhan Moosavi**
The Ohio State University
moosavinejaddaryakenari.1@osu.edu

**Rajiv Ramnath**
The Ohio State University
ramnath.6@osu.edu

**Srinivasan Parthasarathy**
The Ohio State University
srini@cse.ohio-state.edu

## Abstract

In this paper, we present a framework for Question Difficulty and Expertise Estimation (QDEE) in Community Question Answering sites (CQAs) such as Yahoo! Answers and Stack Overflow, which tackles a fundamental challenge in crowdsourcing: how to appropriately route and assign questions to users with the suitable expertise. This problem domain has been the subject of much research and includes both language-agnostic as well as language conscious solutions. We bring to bear a key language-agnostic insight: that users gain expertise and therefore tend to ask as well as answer more difficult questions over time. We use this insight within the popular competition (directed) graph model to estimate question difficulty and user expertise by identifying key hierarchical structure within said model. An important and novel contribution here is the application of "social agony" to this problem domain. Difficulty levels of newly posted questions (the cold-start problem) are estimated by using our QDEE framework and additional textual features. We also propose a model to route newly posted questions to appropriate users based on the difficulty level of the question and the expertise of the user. Extensive experiments on real world CQAs such as Yahoo! Answers and Stack Overflow data demonstrate the improved efficacy of our approach over contemporary state-of-the-art models.

## Introduction

Community question answering systems (CQAs) such as Stack Overflow and Yahoo! Answers are examples of social media sites, with their usage being examples of an important type of computer supported cooperative work in practice. In recent years, the usage of CQAs has seen a dramatic increase in both the frequency of questions posted and general user activity. This, in turn, has given rise to several interesting problems ranging from expertise estimation to question difficulty estimation, and from automated question routing to incentive mechanism design on such collaborative websites (Shen et al. 2015; Fang et al. 2016; Rieh et al. 2017). For example, Liu et al. (Liu, Song, and Lin 2011; Liu et al. 2013) introduced a two-player and no-draw version of TrueSkill (Herbrich, Minka, and Graepel 2007) to estimate question difficulty level in CQAs. They proposed a competition-based approach which takes into account both question difficulty and user expertise. Both of these approaches are simple and *language agnostic* but suffer from an inherent data sparseness problem (limited number of relationships between questions and users). To alleviate the data sparseness problem, Wang et al. (Wang, J. Liu, and Guo 2014) and Pal (Pal 2015) proposed *language-conscious* methods that exploit the textual descriptions (e.g., titles, body, and tags) of the questions to do question difficulty estimation (QDE), based on the assumption that if two questions are close in their textual descriptions, they will also be close in their difficulty levels.

A key observation we make in this paper is that there is an inherent monotonicity-like characteristic to the expertise-level of a user and the difficulty level of questions posed by the same user within such CQAs.

Table 1: Example of questions asked by a Stack Overflow user in Python. The difficulty levels of questions are increasing over time.

| Questions in Python | Question-Date |
|---|---|
| $q_1$: use basic build-in function *sum* on a *list* | July 2013 |
| $q_2$: changing a list element in multiple lists | Sept. 2013 |
| $q_3$: *list comprehension* and *generator* | Oct. 2013 |
| $q_4$: copying 2-D Python list of arbitrary length | Feb. 2014 |
| $q_5$: using *regular expressions* in *Python* | Nov. 2014 |

To illustrate this characteristic we provide an example of a real (but anonymized) user $u$ in Stack Overflow, who asked several questions about *Python*, as shown in Table 1. Two experts of *Python* were asked to provide a ranking order of these questions based on their difficulty levels. They both agreed that the difficulty level ranking order is: $q_1 \prec q_2 \prec q_3 \prec q_4 \prec q_5$ (i.e., in ascending difficulty level). We also observed that $u$ gained expertise over time, as shown by the answers that $u$ provided. For example, $u$ provided an answer $a_1$ to a question about "installing *new regex module* in *Python*" on March 21, 2015, after $u$ gained experience in *regular expressions* from asking and receiving answers to question $q_5$, thus becoming capable of providing an answer like $a_1$. The example above highlights the fact that as a user's expertise increases, he or she is capable of both asking as well as answering harder questions. We refer to this observation as the *Expertise Gain Assumption* (EGA) and note that it is *language agnostic*. This observation is consistent with an analysis of 18 interviews by Rieh et al. (Rieh et al. 2017), where they reported that as a result of providing answers in a CQA environment, users (teens) were able to expand their knowledge bases, improve inquisitiveness, embrace challenging questions, and increase self-confidence. Moreover, a recent study on quality of content in a CQA environment by Ferrara et al. (Ferrara et al. 2017), revealed a long-term trend in terms of improvement of quality of answers due to learning and skill acquisition by users over time.

The above idea can be used to build a directed competition graph, in which edges are drawn from vertices with lower question difficulty levels (or user expertise scores) to vertices with higher question difficulty levels (or user expertise scores). This competition graph exhibits a strong hierarchical structure. In the ideal case, if the competition graph were acyclic, vertices can be partitioned into a natural hierarchy such that there are only edges from lower difficulty levels (or expertise scores) to higher difficulty levels (or expertise scores). However, in practice, competition graphs are not acyclic; thus we need to penalize the edges that violate the hierarchy. After examining several ideas, we empirically found that *social agony* (Gupte et al. 2011; Tatti 2015; Sun et al. 2017) works well for this problem. Social agony is a measure where each edge that violates the hierarchy is penalized based on the severity of the violation. We note that the *application of social agony to the question difficulty and expertise estimation problem in collaborative question answering systems has not been considered previously*.

The above approaches can estimate the difficulty level of resolved questions, but cannot solve the *cold-start problem*, that is, estimating the difficulty level of newly posted questions (which naturally have limited context). To address the cold-start problem, we leverage a combination of EGA and language cognizant features to provide the necessary context for the difficulty estimation procedure. We also then devise a strategy to identify a candidate set of potential users that the question may be routed to (i.e., potential answerers). We lever the prior history of the user, the estimated expertise provided by our framework, and (optional) textual features to rank users within the candidate set.

We evaluate our approach on several real-world CQAs such as Stack Overflow and Yahoo! Answers (Japanese). We find that our insight of explicitly modeling expertise and question difficulty as a monotonically increasing function in our competition graph model, results in significant improvement of question difficulty estimation (when compared with the state-of-the-art competition models and other baselines). We also find that the use of social agony, a new idea in the domain of question difficulty estimation, outperforms other approaches for approximating the hierarchical structure of the resulting competition graph. Additionally, we find that understanding user expertise and accounting for question difficulty estimation simultaneously, yields interesting insights on the real-world datasets that we used for evaluation.

## Related Work

A two-player and no-draw version of TrueSkill (Herbrich, Minka, and Graepel 2007) was introduced in (Liu, Song, and Lin 2011; Liu et al. 2013) by Liu et al. to estimate question difficulty in CQAs. The result of learning skills in the competition game for question nodes is their *difficulty* score and for user nodes is their *expertise* score. The main shortcomings of TrueSkill are its sensitivity to the false-positive cases[1] and also the tendency to overfit. We discuss more about these problems in Experiments section.

Hanrahan et al. (Hanrahan, Convertino, and Nelson 2012) proposed to develop indicators for hard problems and experts. They assumed that questions which took longer to receive their best answer, require a higher degree of expertise to be resolved. Huna et al. (Huna, Srba, and Bielikova 2016) leveraged this assumption and proposed to calculate a user's reputation based on the difficulty of the questions which he or she asked. Yang et al. (Yang et al. 2014) proposed that harder questions can generate more answers or discussions than easier ones. They called the number of answers provided for a question as debatableness, which is a very important factor for determining the expertise of users in their model. However, Yang et al. did not provide any evidence to support their assumption, since they lacked the information of question difficulty. In this paper, we conducted some experiments to verify this assumption by using the question difficulty scores estimated by our model. It is worth mentioning that neither of assumptions by Hanrahan et al. (Hanrahan, Convertino, and Nelson 2012) (which is also used in (Huna, Srba, and Bielikova 2016)) nor Yang et al. (Yang et al. 2014) can be leveraged to estimate difficulty level of newly posted questions, since such questions have no answers.

Wang et al. (Wang, J. Liu, and Guo 2014) proposed a language-conscious solution that uses a regularized competition model (RCM), which formalizes the QDE process as one that tries to minimize a loss on question-user comparisons with manifold regularization on questions' textual descriptions. They add additional context by assuming that difficulty levels of newly posted questions are similar to that of questions with similar textual descriptions. Similarly, Pal (Pal 2015) formalizes QDE to a convex optimization problem. Both of these approaches are language-conscious and do not easily scale to large competition graphs (as a point of comparison, we scale to graphs with a larger order of magnitude).

To summarize, we propose an approach to avoid overfitting, to address the cold-start problem, and also to improve the scalability of the solution.

## Problem Statement

Assume we are given three datasets of Questions $Q = \langle Q_1, Q_2, \ldots, Q_N \rangle$, Answers $A = \langle A_1, A_2, \ldots, A_M \rangle$, and Users $U = \langle U_1, U_2, \ldots, U_K \rangle$. For each question $Q_i \in Q$, we have a tuple of the form $\langle Asker_i, Answers_i, BestAnswer_i, Score_i, Prize_i, Category_i \rangle$, where $Asker_i \in U$, $Answers_i \subseteq A$, $BestAnswer_i \in A$, $Score_i$ is an integer, $Prize_i$[2] is a non-negative integer, and $Category_i$ is a set of keywords which $Q_i$ belongs to. Additionally, for each answer $A_i \in A$, we have a tuple of the form $\langle Answerer_i, Score_i \rangle$, where $Answerer_i \in U$ and $Score_i$ is an integer. Given the preliminaries, in this work we focus on the following problems:

- Question Difficulty Estimation: given a set of questions $\widehat{Q} = \langle \hat{q}_1, \hat{q}_2, \ldots, \hat{q}_n \rangle \subseteq Q$ which belong to a category $t$ (i.e., we have: $\{\forall \hat{q} \in \widehat{Q} | t \in \hat{q}.Category\}$), the goal is to propose a ranking function $f$ which can sort questions in category $t$ based on their difficulty level. By using the function $f$, we expect to see relations like $\hat{q}_i \prec \hat{q}_j$, $1 \leq i, j \leq n$, which means question $i$ is easier than question $j$.
- Expertise Estimation: given a set of users $\widehat{U} = \langle \hat{u}_1, \hat{u}_2, \ldots, \hat{u}_n \rangle \subseteq U$, who have contributed to resolve at least one question in category $t$, the goal is to learn a function $g_t : \hat{u} \in \widehat{U} \longrightarrow R$, that returns an expertise score $s \in R$ for each user $\hat{u} \in \widehat{U}$, based on questions in category $t$, and their related difficulty levels.
- Question Routing: given a question $q \in Q$ of category $t$, we look for an ordered list of experts $(U_{a_1}, U_{a_2}, \ldots, U_{a_\kappa})$, $U_{a_i} \in U$ for $1 \leq i \leq \kappa$, who are the top-$\kappa$ potential resolvers for question $q$ based on their expertise in category $t$ of questions.

---
[1]In a subjective case, like when an asker is asking about the opinion of the community about a topic, the expertise of the asker is not necessarily less than the difficulty of the question or expertise of the answerers.

[2]Sometimes, an asker may motivate other users for contribution by defining some prize to be assigned to the best answerer.

## Methodology

In this section, we describe different components of the proposed Question Difficulty and Expertise Estimation (QDEE) framework. Figure 1 shows the overall process of the QDEE framework. The key steps in this process are: 1) Building the competition graph to incorporate our intuition (monotonicity constraints); 2) Examining the use of different heuristic-based algorithms to estimate question difficulty; 3) Leveraging *EGA* and language cognizant features to estimate difficulty levels for newly posted questions; and 4) Routing newly posted questions to potential answerers, identified by using textual features, questions difficulty level, and user expertise rank generated by previous steps of QDEE.

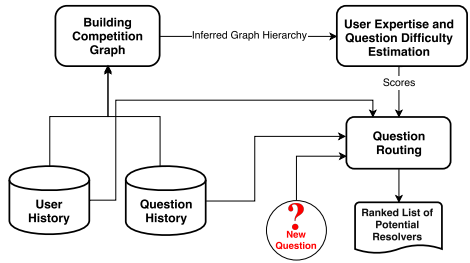

Figure 1: The overall process of the QDEE framework

### Building the Competition Graph

In CQAs, when an asker $u_a$ posts a question $q$, there will be several answerers who answer $q$. One answer will typically be selected as the best answer by the asker $u_a$ or voted by the community. The user who provides the best answer is named as the best answerer $u_b$ and the set of non-best answerers are denoted as $S = \{u_{o1}, u_{o2}, \ldots, u_{om}\}$. In a recent work, Liu et al. (Liu et al. 2013) made two assumptions when building a competition graph model which are: 1) Given a question answering thread, the difficulty score of the question $q$ is higher than the expertise score of the asker $u_a$, but lower than that of the best answerer $u_b$, and 2) The expertise score of the best answerer $u_b$ is higher than that of the asker $u_a$ as well as any answerer in the non-best answerers set $S$. The difficulty score of question $q$ is not assumed to be lower than the expertise score of any answerer in $S$, since such a user may just happened to see the question and responded that, rather than knowing the answer well (Wang, J. Liu, and Guo 2014). Take the category Python in Stack Overflow for example, it is common to have answers like "method $x$ provided by user $a$ works for Python 2.7, but I have trouble in running it with Python 3.0". These kinds of answers do not show answerers' expertise are higher than questions' difficulty levels.

Taking a competitive viewpoint, each pairwise competition can be viewed as a two-player competition with one winner and one loser. So, edges in the competition graph are generated as follows, where question $q$ can be viewed as a pseudo user $u_q$:

- from asker $u_a$ to question $q$: $(u_a, q)$
- from question $q$ to the best answerer $u_b$: $(q, u_b)$
- from asker $u_a$ to the best answerer $u_b$: $(u_a, u_b)$
- from each answerer in the set of non-best answerers $S$ to the best answerer $u_b$: $(u_o, u_b), \forall u_o \in S$

However, this competition graph suffers from the following data sparseness problem. Each question has only one in-edge (from asker) and one out-edge (to the best answerer), which might not provide enough information to achieve an accurate estimation. One observation we seek to leverage to combat this sparseness problem is the fact that users typically gain expertise across multiple interactions with the CQA and tend to ask more difficult questions within the same domain over time, which is referred *EGA* as we shown in Table 1.

We formalize the *EGA* as follows: suppose user $u_a$ asked questions $(q_0, q_1, \ldots, q_l)$. If $t(q_i) + \tau \leq t(q_{i+1})$, $\forall i \in [0, l-1]$, which means that question $q_{i+1}$ was asked after $q_i$ and the time interval is bigger than a threshold $\tau > 0$, we then consider that difficulty score of $q_i$ is lower than $q_{i+1}$ and an edge $(q_i, q_{i+1})$ will be added to the competition graph. Additional edges such as $(q_i, q_j)$ are added to alleviate the data spareness problem, where $\forall i, j \in [0, l]$ and $t(q_i) + \tau \leq t(q_j)$, and the goal being to improve the inference on question difficulty estimation.

Figure 2 illustrates this process. A user $u_1$ in Stack Overflow asked three questions $q_1$, $q_2$, and $q_3$ in Jan. 2014, Aug. 2014 and Dec. 2014, respectively. These three questions are from the same domain *Python*. In addition to edges $(u_1, q_1)$, $(u_1, q_2)$, and $(u_1, q_3)$, three blue edges $(q_1, q_2)$, $(q_2, q_3)$ and $(q_1, q_3)$ are added into the competition graph by leveraging *EGA*. Users $u_2$, $u_3$ and $u_4$ provided answers for question $q_3$, and $u_4$ was selected as the best answerer. These activities generate edges: $(u_1, u_4)$, $(q_3, u_4)$, $(u_2, u_4)$, and $(u_3, u_4)$. Then user $u_4$ asked question $q_4$ in Jan. 2015, and $q_4$'s best answerer is $u_1$. Above activities generate edges $(u_4, q_4)$ and $(u_4, u_1)$, and $(q_4, u_1)$. It is important to note that based on this real-world example, cycles are inevitable in the generated competition graph.

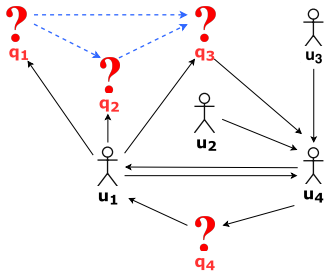

Figure 2: Competition graph illustration: blue dashed links $(q_1, q_2)$, $(q_1, q_3)$, and $(q_2, q_3)$ are edges added based on *EGA*, which aim to solve data sparseness problem. Other links are built based on previous work (Liu et al. 2013; Wang, J. Liu, and Guo 2014).

### Inferring Graph Hierarchy for Estimation

We next consider the problem of question difficulty estimation in the context of directed competition graphs: given a directed competition graph, partitioning vertices into the ranked groups (or equivalence classes) such that there are only edges from lower groups (vertices with lower difficulty level or expertise score) to upper groups (vertices with higher difficulty level or expertise score). However, such a perfect partitioning is only possible if the competition graph is a directed acyclic graph (DAG). In such a case, the topological ordering of the DAG can be used to infer question difficulty and user expertise. Since cycles are inherently present in graphs induced from such CQAs (see real-world example in Figure 2), next we examine strategies to reduce the competition graph to a DAG.

**Reducing the Competition Graph to DAG:** Reducing a general directed graph to a DAG in an optimal fashion is the celebrated *minimum feedback arc set problem*, a well-known NP-hard problem. All known heuristic methods must obey the fact that the minimum feedback arc set problem is approximation resistant. It means that, in practice, the difference between the solution found by a heuristic and the optimal solution can be as large as $O(n)$. Note also that strategies to detect communities in directed graphs may not be useful since they optimize a different criterion function(Shih et al. 2014; Satuluri and Parthasarathy 2011).

**DFS Heuristic:** A simple, fast, and domain independent approach for eliminating cycles is to leverage depth-first search (DFS) elimination. We can perform a DFS starting from bottom vertices (zero in-degree) in the competition graph. A back edge pointing back to a node closer to the bottom of the graph is deleted, as it violates the hierarchy. The DFS based approach cannot guarantee optimality - it cannot ensure that the links ignored during the graph traversal in order to prevent loops from happening are actually the appropriate edges to be removed. Which edge will be ignored, solely depends on the order in which the graph is traversed.

Another way to reason about this problem is to define a penalty function $p$ on the edges. Edges which violate the hierarchy (from a lower group to a upper group) can be penalized. Given a penalty function, the task is to find the hierarchy that minimizes the total penalty (sum of penalties of all edges in the competition graph). Questions' difficulty level can then be inferred from the graph's hierarchy information. There are many choices for the penalty function. The most simplest way is to define a constant penalty for any edge that violates the hierarchy. If the penalty is set as a constant 1, then this problem is equivalent to a minimum feedback arc set problem as we discussed earlier.

**Agony Heuristic:** A more practical variant is to penalize the violating edges by the severity of their violation, which means that edges that respect the hierarchy receive a penalty of 0 and penalty increase linearly as the violation becomes more severe. This particular approach is referred to as social agony (Gupte et al. 2011). Given a network $G = (V, E)$ which contains cycles, each node $v$ has a rank $r(v)$. Higher rank nodes are less likely to connect to lower rank nodes. Hence, directed edges that go from higher rank nodes are less prevalent than edges that go in reverse direction. If $r(u) > r(v)$, then edge $u \Rightarrow v$ causes agony to the user $u$ and the amount of agony depends on the difference between their ranks. Gupte et al. (Gupte et al. 2011) defined the agony to $u$ caused by edge $(u, v)$ is equal to $max(r(u) - r(v) + 1, 0)$. Since nodes typically minimize their agony, the problem is changed to find a ranking $r$ that minimize the total agony in the graph. Sun et al. (Sun et al. 2017) proposed to infer graph hierarchy using a range of features, including TrueSkill and social agony. They also devised several strategies to leverage the inferred hierarchy for removing a small subset of edges to make the graph acyclic.

The ranking scores in the resulting ranking $r$ of question nodes in the graph are question difficulty scores, and the learned skills of all other users can be thought to correspond to their expertise scores. We note that the astute reader may have noted that, on the one hand, our EGA assumption adds edges, while, the Agony heuristic removes edges to create a DAG-structure. Intuitively, a large fraction of edges we add (EGA) are important constraints (that are not removed by the agony heuristic), and finally lead to improved estimates on question difficulty as we shall demonstrate in our empirical evaluation.

### Cold-Start Estimation and Routing

We have thus far discussed how to estimate difficulty levels of resolved questions, from which competition edges could be extracted. However, such an approach cannot address estimating the difficulty of newly posted questions without any answers received. This challenging problem is the *cold-start problem*, sometimes referred to as the "item cold-start problem" (Chang, Harper, and Terveen 2015) in recommender systems (Wang et al. 2012; Sun et al. 2012; Wang et al. 2014).

**Cold-Start Difficulty Estimation:** Wang et al. (Wang, J. Liu, and Guo 2014) applied a text-based $K$-Nearest Neighbor (KNN) approach to cold-start estimation in CQAs. Given a cold question $q^*$, $k$ well-resolved questions that are closest to $q^*$ in textual descriptions, are picked as its nearest neighbors. The difficulty score of $q^*$ is predicted as the averaged difficulty scores of its nearest neighbors. The authors employed a Boolean term weighting schema to represent a cold-start question, and subsequently levered the Jaccard Coefficient to select its nearest neighbors. $q^*$'s predicted difficulty level by

KNN approach can be represented as $d_{knn}(q^*)$.

Our idea is to enhance the above approach by leveraging the *EGA* which we believe can bridge the gap between cold-start and well-resolved questions asked by the same user. Suppose the most recent $k$ questions asked by user $u$ are $q_1, q_2, ..., q_k$ and their associated difficulty levels are available to us (estimated by QDEE). Then, $q^*$'s difficulty level ($d_{ega}(q^*)$) can be inferred from these most recent-$k$ questions by several strategies such as:

- Minimum difficulty level of $q_1, ..., q_k$, represented as *Min*
- Maximum difficulty level of $q_1, ..., q_k$, represented as *Max*
- Average difficulty level of $q_1, ..., q_k$, represented as *Avg*
- Difficulty level of the most recently posed question, which is $q_k$, represented as *MRQ*

We can then combine these two approaches, and the difficulty level of the cold question $q^*$ estimated by the hybrid model is:

$$d(q^*) = \alpha \cdot d_{knn}(q^*) + (1-\alpha) \cdot d_{ega}(q^*) \qquad (1)$$

where $\alpha \in [0,1]$, is a simple regularization parameter. We note that it is possible that the user posing the question is a new user (or one that has not posted a sufficient number of questions). In this case, we simply set $\alpha = 1$.

**Identifying a Candidate Set of Potential Answerers:** To identify potential users to route new questions to, we need to identify a candidate set of users. We rely on a similar procedure (as we call that **QT**) to what we have outlined above – we compute a set $Set_T$ of users who have attempted to answer questions that are textually similar to the current question being posed. We also identify a set of answerers of questions of similar difficulty to the current question within the domain. Since we have a ranked list of questions (based on their question difficulty score – from the competition graph induced from the training data), we simply identify a group of questions with similar or slightly greater difficulty and compute the union of all their answerers (we refer to this set as $Set_Q$). We then select $\beta \cdot k$ and $(1-\beta) \cdot k$ users with top activity level[3] and user expertise in $Set_T$ and $Set_Q$, respectively, where $\beta \in [0,1]$.

The union of $Set_T \cup Set_Q$ represents our candidate set of answerers $Set_C$. We then refine this set by ranking elements of this set according to their estimated expertise from QDEE and user activity levels. Questions can then be routed to potential answerers according to this ranked list.

There are two variants of the algorithm **QT**:

- A language conscious model, represented as **T**: Candidate set $Set_C$ is generated by selecting $k$ users with top user activity level and expertise from $Set_T$, without using set $Set_Q$ ($\beta = 1$).
- A language agnostic algorithm, named as **Q**: Candidate set $Set_C$ consists of $k$ users who have top user activity level and estimated user expertise from QDEE, without using set $Set_T$ ($\beta = 0$).

---

[3] A user $u$'s activity level is defined as $(1 + \frac{\text{\# questions u has answered}}{n})$, where $n$ is used for normalization.

## Characterizing Potential Answerers

In the previous section we have described a methodology for simultaneously estimating question difficulty and user expertise in CQAs for pre-specified domains. We also described an approach to address the challenging cold-start problem that arises in such CQAs. In this section we briefly describe some ideas that leverage question difficulty estimation to understand the different categories of users that actively participate in such systems.

We hypothesize that users in CQAs can be characterized by their activity and effectiveness on a range of questions. The framework described in the previous section allows us to drill down further on this aspect. One can extract a set of features associated with each user, where some examples can be "how often the user answers questions of varying difficulty (e.g. easy, medium, and hard)", "how often said user is deemed as the best answerer of each category of question (e.g. easy, medium, and hard)", or some other statistics about the activity like number of logins, number of questions answered, etc.

One can then take such feature vectors for each user and cluster users by levering an appropriate approach. While we are not wed to a particular approach in this article, a simple k-means algorithm with a suitable distance measure or an approach based on Non-negative Matrix Factorization (NMF) (Cheng et al. 2017) are options that one can explore to understand and characterize active participants in such CQAs. For example, one may find some users who attempt to answer all range of questions, however, they are never rated particularly highly (enthusiastic participants) or others who focus only on effectively answering the hardest questions (focused expert) and yet others who enjoy effectively answering all-comers (all-round experts).

These users' patterns of answering questions can be beneficial to filtering candidate set of potential answerers for a newly posted question. For example, a newly posted question $q$ is estimated by our model as a hard question, and users in set $H$ are users who answer hard questions effectively. The candidate set of potential answerers for $q$ is $Set_{C_q}$ (see Methodology section for more details). Then, the filtered candidate set of potential answerers will be $Set_{C_q} \cap H$.

## Experiments

In this section, we evaluate the proposed QDEE framework. We begin by describing the experimental settings (datasets, measures-of-interest, etc.) and then providing detailed analysis for each part of the framework.

### Experimental Settings

The first step is to describe the CQA datasets which we use for evaluation of our QDEE framework.

**Datasets:** We use the Yahoo! Chiebukuro (Japanese Yahoo! Answers) database (2nd edition)[4], and the Stack Overflow as the main datasets for our experiments. The Yahoo! Chiebukuro dataset roughly represents 5 years of data (from April 7, 2009 to April 1, 2014) and contains $16,257,422$ questions and 50,053,894 answers. There are 16 categories provided in the data set. Each question belongs to exactly one category.

---

[4] Yahoo! Chiebukuro Data (2nd edition) provided by National Institute of Informatics by Yahoo Japan Corporation.

The Stack Overflow dataset that we use in this paper is a subset of the one of the recent data dump of Stack Overflow[5], and covers several important tags of questions (e.g. *Java*, *Python*, *C#*, *PHP*, and *HTML*)[6]. The questions in the Stack Overflow dataset represent data for over eight years (from July 2008 to March 2017). More details about the Stack Overflow dataset can be found in Table 2, where $Q$ is short for questions and $U$ is short for users.

Table 2: Stack Overflow dataset for QDE

| Tag | #Q | #Answers | #Unique U | #Q With Bounty |
|---|---|---|---|---|
| Java | 547,509 | 1,302,619 | 359,559 | 11,940 |
| Python | 388,725 | 737,033 | 210,242 | 5,977 |
| C# | 619,174 | 1,302,566 | 271,949 | 11,594 |
| PHP | 508,251 | 1,022,306 | 296,441 | 7,838 |
| HTML | 356,446 | 763,831 | 306,806 | 4,188 |

**Ground Truth:** To evaluate the question difficulty estimation on above mentioned datasets, we leverage the notion of *coin* in the Yahoo! Chiebukuro database (ranging from 0 to 500) and *bounty* in Stack Overflow (ranging from 0 to 550). These coins (bounties) are given to the user who provides the best answer. The prize is specified at the time of submission of the question. In Table 2, we can see the number of questions with non-zero bounty score for each of the five popular categories of questions in Stack Overflow. We note that specifically for these five categories, we additionally relied on three subject matter experts (and teachers of the subject matter) to skim through a sample of the questions, and their evaluation almost exactly agrees (agreement ratio of 0.98) with the difficulty ranking obtained by comparing bounty scores. Given two questions $q_1$ and $q_2$, which were provided number of coins (bounties) $c_1$ and $c_2$ ($c_1 \ne c_2$ and $c_1 > 0, c_2 > 0$) to their corresponding best answerers, respectively, we assume that $q_1$ is relatively harder than $q_2$, if $c_1 > c_2$, and vice verse. Based on this assumption, we select questions which were provided non-zero coins as our ground truth for difficulty estimation. The ground truth data is not leveraged intrinsically by the QDEE estimation process and is only used during evaluation.

**Accuracy Metric:** Assume we have a list of question pairs like $\langle (q_1, q_2), (q_1, q_3), \ldots, (q_2, q_3), (q_2, q_4), \ldots (q_{n-1}, q_n) \rangle$, where $q_i$ is in ground truth data for $1 \le i \le n$, that is, we have the value of coin (in case of Yahoo! Answer) or bounty (in case of Stack Overflow) for $q_i$. Thus, for each pair of $(q_i, q_j)$, $1 \le i, j \le n$, we know about their relative difficulty order. In this way, we use a standard evaluation metric, accuracy (*Acc*), for measuring the effectiveness of the QDEE framework, as previous studies (Liu, Song, and Lin 2011; Liu et al. 2013; Wang, J. Liu, and Guo 2014) used that as

---
[5]We used the data dump which is released on March 14, 2017 and is available online at https://archive.org/details/stackexchange

[6]Real-time question frequency report per tags can be find at http://stackoverflow.com/tags

Table 3: Number of Valid Question Pairs in Ground Truth Set

| **Stack Overflow** | Java | Python | C# | PHP | HTML |
|---|---|---|---|---|---|
| # valid Q pairs | 2,923 | 1,969 | 4,827 | 1,745 | 1,823 |
| **Yahoo!** | Manners | News | School | Business | Travel |
| # valid Q pairs | 4,098 | 5,614 | 12,241 | 8,523 | 12,856 |

well. Accuracy is described as follows:

$$Accuracy = \frac{\text{\# correctly predicted question pairs}}{\text{\# valid question pairs}} \quad (2)$$

A question pair is regarded as correctly predicted if the relative difficulty ranking given by an estimation method is consistent with that given by the ground truth. The higher the accuracy is, the better a method performs.

### Difficulty Estimation for Resolved Questions

We next describe our experiments on Question difficulty estimation. Before we do so, we briefly define the baselines that we compare our approach with.

**Baselines:** *DFS-based* (as described in Methodology section) and *TrueSkill-based* (Wang, J. Liu, and Guo 2014; Liu et al. 2013) approaches are used as our baselines. Since others have shown that TrueSkill-based approach outperforms the PageRank-based approach (Wang, J. Liu, and Guo 2014; Liu et al. 2013) (using the relative importance of nodes to estimate question difficulty scores and user expertise scores), here we do not include a comparison with PageRank-based approach. For this part of the evaluation, we focus on language-agnostic methods.

As described earlier, we use accuracy to report the performance of different approaches on resolved questions. Table 3 shows the number of valid question pairs in ground truth for evaluation. Figures 3a and 3b demonstrate the evaluation results for Yahoo! and Stack Overflow dataset respectively. In these figures, suffix "-EGA" means leveraging the *expertise gain* assumption for creating the competition graph (see Building the Competition Graph) [7]. Here, we report the results across five different categories of questions in Yahoo! and Stack Overflow (in the interests of space - results on other sub-domains show similar trends). Based on the results, we can point to several important observations:

**1)** Leveraging the *expertise gain* assumption (EGA) to handle graph sparseness, can significantly improve the question difficulty estimation results (not just for our methods but for the baselines as well). We can observe this improvement on both datasets and different categories of questions. For instance, we can observe the accuracy by TrueSkill is improved by about 5% on average, when we apply the EGA on Yahoo! dataset. We conducted McNemar's test[8] for each category in both datasets to test whether EGA has significant effect or not. Based on the test results, all p-values are well below 0.01%. Hence, these results represent a significant gain by leveraging EGA.

---
[7]The parameter $\tau$ is set as 30 days

[8]The McNemar test, introduced by Quinn McNemar in 1947, is applied to a $2 \times 2$ contingency table, which tabulates the outcomes of two tests on question pairs. A question pair is considered as a positive case if its relative difficulty level is predicted correctly, otherwise negative.

2) By applying the Social Agony for the first time, we obtained better performance results in comparison with existing baselines. As an important observation, we have about 4% improvement on average accuracy for both Yahoo! Answers and Stack Overflow datasets by Social Agony in comparison with TrueSkill. Again we conducted McNemar's test for each category in Yahoo! answers and Stack Overflow dataset, and found that all *p*-values are less than 0.01%, which supports the conclusion that our proposed Social Agony model does lead to significant improvements over the baselines.

3) Empirically, we observe that Social Agony is more robust than TrueSkill - in particular for subjective categories. Both approaches achieve similar performance within technical categories such as *Python*, *Java*, and *C#* in Stack Overflow. These questions largely tend to be rational and objective. However, in general categories, such as *Travel*, *Business*, and *School* within the Yahoo! dataset, questions and answers, which may be designed to provoke discussion, controversy, humor or other emotional responses, tend to be less rational and more subjective. This in turn leads to noise – typically false positive edges that do not necessarily represent the true relationships between question difficulty and user expertise. Manual inspection suggests this is uncommon in the technical categories (e.g. Python) as noted above. TrueSkill is more sensitive to these false positive noisy edges and hence has much worse performance than Social Agony in such general categories. More specifically, TrueSkill tends to overfit – requiring significant updates for ranking scores when it internally accounts for such noisy edges.

We also compared *Agony-EGA* with the work of Hanrahan et al. (Hanrahan, Convertino, and Nelson 2012), Huna et al. (Huna, Srba, and Bielikova 2016) and Yang et al. (Yang et al. 2014), as shown in Figure 3c, even though their question difficulty estimation model cannot be applied to cold-start questions. *Number-Of-Answers* is the approach to estimate question difficulty by using the number of answers provided for the target question (Yang et al. 2014), and *Time-First-Answer* is the model to estimate question difficulty by measuring how long it takes such that the target question gets its first answer (Hanrahan, Convertino, and Nelson 2012; Huna, Srba, and Bielikova 2016). We also consider a modification of *Time-First-Answer* represented as *Time-Best-Answer*, which assumes that a harder question takes longer time to get its best answer than easier ones. Two interesting observations can be concluded as follows:

- *Time-Best-Answer* is a better indicator for question difficulty estimation than *Time-First-Answer*, since active users may post some quick (with un-matching expertise) answers to gain their reputation when they find the target question has no answers so far. However, we have a significant number of questions in Stack Overflow which do not have any best answer specified by the askers.
- *Number-Of-Answer* performs better than *Time-First-Answer* and random guess (50% accuracy). The *Number-Of-Answer* approach is a key factor in the model proposed by Yang's et al. (Yang et al. 2014) model to identify expert users. Intuitively, it assumes that harder questions can generate more answers or discussions than easier ones. By leveraging the difficulty level estimated by our model, we can provide empirical evidence to support this assumption.

**Difficulty Estimation for Cold-Start Questions**

We next examine the question of how effective the QDEE framework is to estimate difficulty of questions for which there is limited context (i.e. the Cold-Start questions).

For this experiment, we split the datasets into the training and testing sets as follows. For a user $u$, we first sort the questions which $u$ asked previously, by date of posting the questions. The first 90% of questions are selected for training, and questions associated with a prize in the remaining 10% of data formed our test dataset. We report results on Stack Overflow[9] for the categories of *C#*, *Java*, and *Python* (in the interests of space - results on other sub-domains show similar trends). Competition edges are extracted only from the training datasets to build competition graphs. Then, the QDEE framework, described in Methodology section, is applied to estimate the difficulty levels of newly posted questions.

**Strategy Selection** We proposed four natural strategies to estimate difficulty level of newly posted questions by leveraging our *EGA*: *Min*, *Max*, *Avg*, and *MRQ* (see detailed description in Methodology section). Figures 4a, 4b, and 4c show accuracies of different strategies for cold-start questions on *Python*, *C#*, and *Java* of Stack Overflow, respectively. We can observe that:

- Strategy *Max* performs better than strategy *Min*, *Avg* and *MRQ* overall, which is consistent with the *EGA*. Users tend to ask more difficult questions, hence it's more reasonable to use the maximum difficulty level of questions asked recently to infer difficulty level of newly posted questions, comparing with other strategies.
- By using the most-recent 3 questions posted by the same asker, strategy *Max* performs the best on all categories.

**Performance of the Hybrid Model** We tested our hybrid model's performances for cold-start questions in *Python*, *Java*, and *C#* of Stack Overflow, with different regularization parameter $\alpha$. When $\alpha = 0$, only $d_{ega}$ is applied to estimate newly posted questions' difficulty levels. In our experiments, we select strategy *Max* and use the most recent 3 posted questions to infer $d_{ega}$. While $\alpha = 1$, only $d_{knn}$ is valid for obtaining difficulty levels of newly posted questions and corresponds to the baseline strategy employed by Wang et al. (Wang, J. Liu, and Guo 2014). After varying $k$, we found the optimal performance at $k = 7$ for *Python*, and $k = 10$ for *Java* and *C#*. The main observations are as follows:

- $d_{ega}$ outperforms $d_{knn}$ by 3%, 7.8%, and 24.9% on *C#*, *Python*, and *Java*, respectively. We conducted McNemar's test for each category, and found that all p-values are less

---

[9]Since the Yahoo! dataset is largely in Japanese and as we do not have any expert in this language for text analysis purposes, we do not report results on the cold-start problem for this dataset.

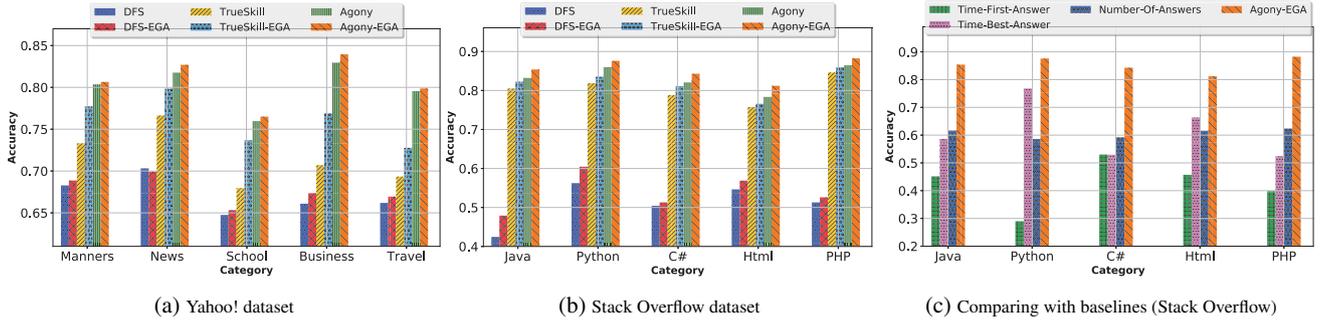

(a) Yahoo! dataset
(b) Stack Overflow dataset
(c) Comparing with baselines (Stack Overflow)

Figure 3: (a,b) The evaluation of question difficulty estimation based on Yahoo! and Stack Overflow datasets. (c) The comparison of proposed solution to estimate question difficulty in Stack Overflow with three baselines *Time-First-Answer* (Huna, Srba, and Bielikova 2016), *Time-Best-Answer* (an adaptation of (Huna, Srba, and Bielikova 2016)), and *Number-of-Answers* (Yang et al. 2014) .

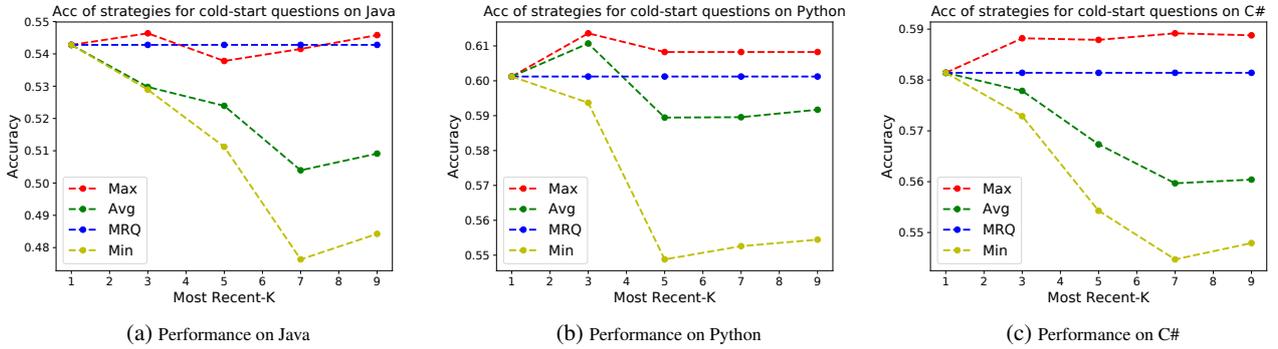

(a) Performance on Java
(b) Performance on Python
(c) Performance on C#

Figure 4: Accuracy of different strategies in $d_{ega}$ for cold-start questions in Stack Overflow.

than 0.01%, which shows the effectiveness of leveraging *EGA* for estimating difficulty levels of cold-start questions.
- By leveraging EGA and textual information of questions, our proposed hybrid model can effectively deal with the cold-start problem. In our experiments, the hybrid model can achieve the best performance when $\alpha = 0.1$. For instance, on *Java*, the hybrid model outperforms $d_{ega}$ and $d_{knn}$ by 8.6% and 35.7%, respectively. Moreover, on *C#*, the hybrid model outperforms $d_{ega}$ and $d_{knn}$ by 4.8% and 7.9%, respectively.

**Performance of Routing Newly Posted Questions** Figure 5 shows the performance of routing newly posted questions to candidate users on Java, C#, and Python category of Stack Overflow. Each question in the test set has a best answerer for evaluation. For a newly posted question $q$ in test set, it has a candidate set of users represented as $C_q$ to route. If $q$'s best answerer is in $C_q$, we say it has a hit in $C_q$. So hit-ratio as $y$-axis is $\frac{\#hits}{size\ of\ test\ set}$. The $x$-axis represents $C_q$'s size. We note that this is a very challenging task since the number of active users in each of categories of our training dataset exceeds several hundred thousands. The recommendation engine is limited to about 1000 users (a small fraction of the total candidate set of users one can route to, as noted in Table 2).

We use two language agnostic methods as base lines.

- **R**: The first baseline is based on Stack Overflow's internal reputation algorithm. In this scheme, we route newly posted questions to users with top reputation scores in Stack Overflow. The reputation score is a rough measurement of how much Stack Overflow trusts a user[10]. The primary way to gain reputation is via activities such as posting good questions and useful answers. Shah et al. (Shah and Pomerantz 2010) demonstrated that the answerer's reputation score is the most significant feature for predicting the best quality answer in the Yahoo! Answers website.
- **E**: The second baseline routes newly posted questions to users with top expertise estimated from our QDEE framework.

Two language conscious options (**T** and **QT**) and one language agnostic (**Q**) include:

- **T:** This strategy leverages textual similarity between the question and nearest neighbor questions (up to a 1000 neighbors) in the training data. It then takes the answerers of this nearest neighbor set and ranks the resulting set of users by expertise level and activity.
- **Q:** This strategy leverages question difficulty estimation from our QDEE framework and identifies nearest neighbor questions (up to a 1000 neighbors) in the training data. It then takes the answerers of this nearest neighbor set and ranks the resulting set of users by activity.
- **QT:** This strategy combines both elements from set $Set_Q$ and set $Set_T$ and rank orders the resulting set of users by expertise level and activity.

[10]http://stackoverflow.com/help/whats-reputation

The results of this experiment is reported in Figure 5 for three categories of Stack Overflow dataset[11]. Based on the results, we can make the following observations:

i. Reputation-based (**R**) and Expertise-only solutions are largely ineffective. While both implicitly or explicitly account for activity, these strategies lack sufficient context to be effective.
ii. Language conscious solutions, **T** and **QT**, are quite effective across all three domains. **QT** outperforms all the other approaches (in one domain by a significant margin).
iii. Among language agnostic solutions, the best strategy is based on question difficulty estimation (**Q**). This strategy performs comparable to **T** on one dataset (C#). Unlike strategies **R** and **E** which always target top-level experts, strategy **Q** which routes questions based on matching user expertise, can help middle-level experts to improve their expertise via collaborations in community. In other words, strategy **Q** utilizes the expertise of the entire community.
iv. Taking users' activity level into consideration for identifying candidate set of users for newly posted questions is almost always effective.
v. Parameter Sensitivity: We also test parameter $\beta$'s sensitivity in **QT**. It shows that when $\beta$ is set between 0.6 and 0.8, **QT** can achieve the best performance.
vi. Overall, the best method achieves a very respectable hit rate of up to 45% – for a very challenging task.

**Characterizing Active Users in CQAs:** We now examine some results that correspond to characterizing potential answerers (see Methodology section). Given a user $u$, suppose questions answered by $u$ are $q_1, q_2, ..., q_l$, and these questions' difficulty levels are estimated by our model as $d(q_1), d(q_2), ..., d(q_l)$. We map $d(q_1), d(q_2), ..., d(q_l)$ to a histogram representation $h_u$. The bin size of $h_u$ is set as 3, corresponding to three difficulty levels: *easy*, *medium*, and *hard*, and each user is represented by distribution of questions' difficulty levels. Suppose $u$ is represented as $h_u = (0.8, 0.15, 0.05)$, which means that 80% of all the questions which $u$ answered are easy, 15% are medium, and 5% are hard questions. Based on $u$'s patterns of answering questions, we can conclude that $u$ prefers to answer easy questions. To automatically discover patterns of users' answering question activities, *k-Means* is applied to group users based on their histogram representations. We filter users to only include users that have answered at least 10 questions. For clustering purpose, we used about $24K$, $23K$, and $11K$ of Stack Overflow users on categories Java, C#, and Python, respectively. We set the number of clusters as 50. We then plot these clusters' centers in Figure 6, where the point size is correlated with cluster size in this figure. Users tend to have strong preferences (see figure caption) ranging from focused experts to all-rounders to beginners. Those users who prefer to answer hard questions (represented as cluster *A* in Figure 6) are classified as *Owls* by Yang et al. (Yang et al. 2014), while users who tend to actively answer easy questions in order to gain their reputation (such as users in cluster *H* in figure 6) are named as *sparrows*. In our experiments, only 3% - 10% of users are considered as *Owls* in Stack Overflow.

## Conclusion and Future Work

In this paper we present QDEE, a framework for simultaneously estimating question difficulty and user expertise in CQAs, which tackles a fundamental challenge in crowdsourcing: how to appropriately route and assign questions to suitable answerers. A central element of our design is the insight that users gain expertise over time (EGA assumption) within a competition graph model framework. Our basic approach is language agnostic and demonstrates the effectiveness of the EGA assumption and social agony on two CQAs with different base languages (English and Japanese). We rely on textual features (to identify semantically similar questions) as well as estimated question difficulty to generate related context, and subsequently use this to estimate difficulty level of newly posed questions and route them to appropriate users.

As extension of current study, we would like to examine mechanisms to scale our approach to larger data. We are also interested in the problem of routing newly posted questions (item cold-start) to newly registered users (user cold-start). Finally, we wish to examine a deeper question in a more collective sense – i.e., how does a community at large (e.g. Python community) gain expertise and how can such insights improve the performance of CQAs.

**Acknowledgments** This work is supported by NSF grants CCF-1645599 and IIS-1550302 and a grant from the Ohio Supercomputer Center (PAS0166).
**Code and Data**: https://github.com/zhenv5/QDEE.

---

[11] For the sake of space, we omit the results for categories PHP and HTML of Stack Overflow.

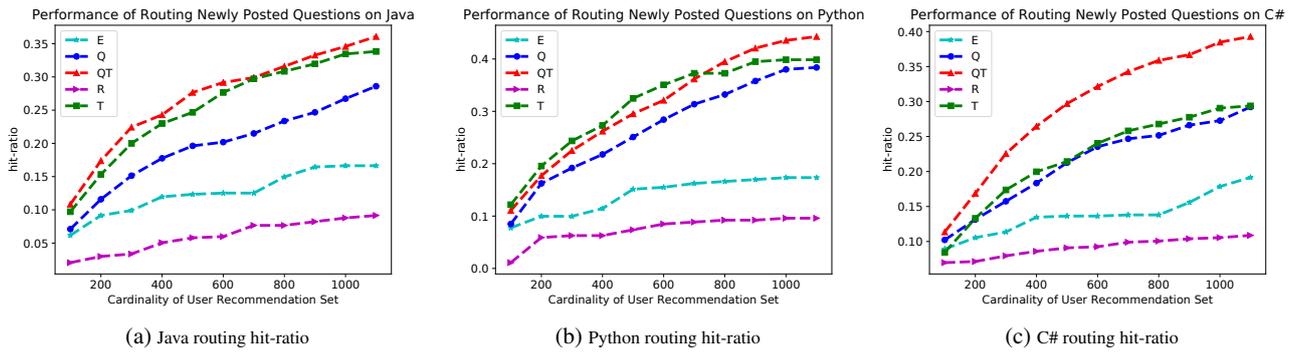

Figure 5: Hit-ratio of routing newly posted questions to users (experts) in Stack Overflow.

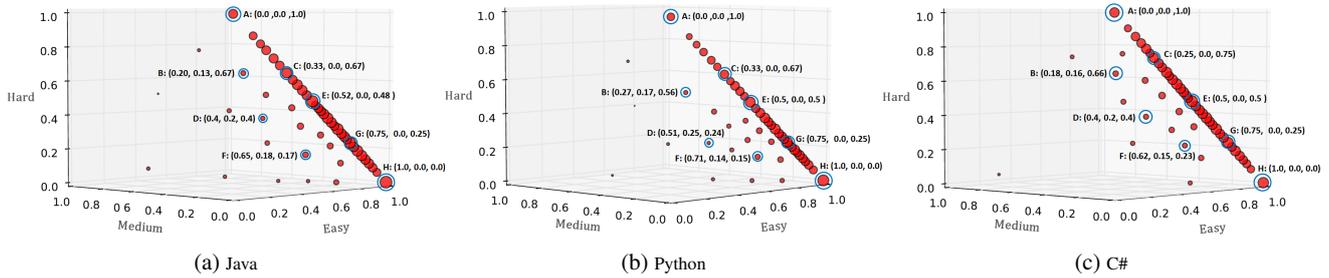

Figure 6: Characterizing Users by their Answering Patterns. Each point in the visualization corresponds to a cluster of users (point size is correlated to cluster size), represented by their mean. Clusters range from Focused Experts (A) to Beginners (H) and a mixture of all-rounders (B,C,D,E,F,G).